\documentclass[%
 reprint,
superscriptaddress,
 amsmath,amssymb,
 aps,
]{revtex4-2}
\usepackage{xcolor}

\usepackage{graphicx}% Include figure files
\usepackage{dcolumn}% Align table columns on decimal point
\usepackage{bm}% bold math
\begin{document}

\preprint{APS/123-QED}

\title{Long-lived population inversion in resonantly driven excitonic antiferromagnet}% Force line breaks with \\

\author{Jacob A. Warshauer}
\author{Huyongqing Chen}
\author{Daniel Alejandro Bustamante Lopez}
\affiliation{Department of Physics, Boston University, 590 Commonwealth Avenue, Boston, MA 02215, USA}
\author{Qishuo Tan}
\affiliation{Department of Chemistry, Boston University, 590 Commonwealth Avenue, Boston, MA 02215, USA}
\author{Jing Tang}
\affiliation{Department of Chemistry, Boston University, 590 Commonwealth Avenue, Boston, MA 02215, USA}
\author{Xi Ling}
\affiliation{Department of Chemistry, Boston University, 590 Commonwealth Avenue, Boston, MA 02215, USA}
\affiliation{Division of Materials Science and Engineering, Boston University, 590 Commonwealth Avenue, Boston, MA 02215, USA}
\affiliation{Photonics Center, Boston University, 8 Saint Mary's St., Boston, MA 02215, USA}
\author{Wanzheng Hu}
\email{wanzheng@bu.edu}
\affiliation{Department of Physics, Boston University, 590 Commonwealth Avenue, Boston, MA 02215, USA}
\affiliation{Division of Materials Science and Engineering, Boston University, 590 Commonwealth Avenue, Boston, MA 02215, USA}
\affiliation{Photonics Center, Boston University, 8 Saint Mary's St., Boston, MA 02215, USA}

\date{\today}% It is always \today, today,

\begin{abstract}
Van der Waals magnets are an emerging material family for investigating light-matter interactions and spin-correlated excitations. Here, we report the discovery of a photo-induced state with a lifetime of 17 ps in the van der Waals antiferromagnet NiPS$_3$, which appears exclusively with resonant pumping at 1.476 eV in the antiferromagnetic state. The long-lived state comes with a negative photoconductivity, a characteristic optical response of population inversion. Our findings demonstrate a promising pathway to  potentially achieve long-lived lasing at terahertz frequencies in reduced dimensions. 
\end{abstract}

\maketitle
Excitons are electron-hole pairs bound by Coulomb attraction that broadly exist in condensed matter systems. Optically driven excitonic systems host a vast range of compelling features, including intraexcitonic resonances\cite{Kaindl2003}, the Hubbard exciton\cite{Mehio2023}, exciton sensing of coherent magnon\cite{Bae2022}, and spin-polarized spatially indirect exciton\cite{Mori2023}. The beauty of rich excitonic physics, however, is eventually limited by the exciton's lifetime before the electron-hole recombination. Long-lived excitons are appealing for the realization of exciton condensation\cite{Eisenstein2014,Kogar2017,Haque2024} and optoelectronic device applications\cite{Mak2016}, making materials with long-lived excitonic states highly sought after.

Van der Waals magnets are an emerging material family for investigating light-matter interactions and spin-correlated excitations\cite{Kang2020,Wang2021,Hwangbo2021, Kim2023, Dirnberger2022,Afanasiev2021,Belvin2021,Ergeçen2022,Mai2021,Lee2016,Tan2022}. The discovery of an ultra-narrow photoluminescence (PL) peak in the antiferromagnetic material NiPS$_3$, identified as a spin-correlated exciton state, opening up novel opportunities to study coherent many-body excitons\cite{Kang2020, Jana, Dirnberger2022, Klaproth2023}. The 330-$\mu$eV linewidth of the exciton peak in NiPS$_3$\cite{Wang2021,Hwangbo2021} is significantly narrower than the typical sub-10 meV exciton linewidth found in two-dimensional semiconductors\cite{Ross2013,Cadiz2017,Wierzbowski2017}, which makes NiPS$_3$ an ideal platform to investigate the exciton dynamics. Apart from the ultra-narrow exciton linewidth, the other fascinating character is the connection between the exciton and magnetic ordering. The exciton PL peak appears only in the antiferromagnetic phase\cite{Wang2021,Hwangbo2021,Kim2023} and displays a splitting under the application of an in-plane magnetic field\cite{Jana, Wang2024}. The exciton peak is highly anisotropic with a temperature dependence following the in-plane magnetic susceptibility anisotropy\cite{Hwangbo2021}, and the exciton dispersion follows closely the double-magnon dispersion\cite{He2024}, indicating a close relationship between the exciton and the magnetic order. Despite extensive interest in the ultra-narrow spin-correlated exciton, there are no studies on resonant driving at exciton levels in NiPS$_3$ or other van der Waals magnets in this family.

Here we use resonant optical excitation to populate the spin-correlated exciton state in the antiferromagnet NiPS$_3$. We probe the charge dynamics using time-resolved terahertz (THz) spectroscopy, which has played a major role in probing intra-exciton transitions\cite{Kaindl2003,Mehio2023,Huber2006}. By tuning the pump photon energy to two of the excitonic features and an energy above the absorption onset, we identify a distinct state tied to the spin-correlated exciton with a lifetime of 17 ps. Further, we show that the long-lived state exhibits a negative photoconductivity over the entire THz probe range, which we attribute to population inversion involving the excitonic ground state.

\begin{figure} 
\includegraphics[scale = 0.55]{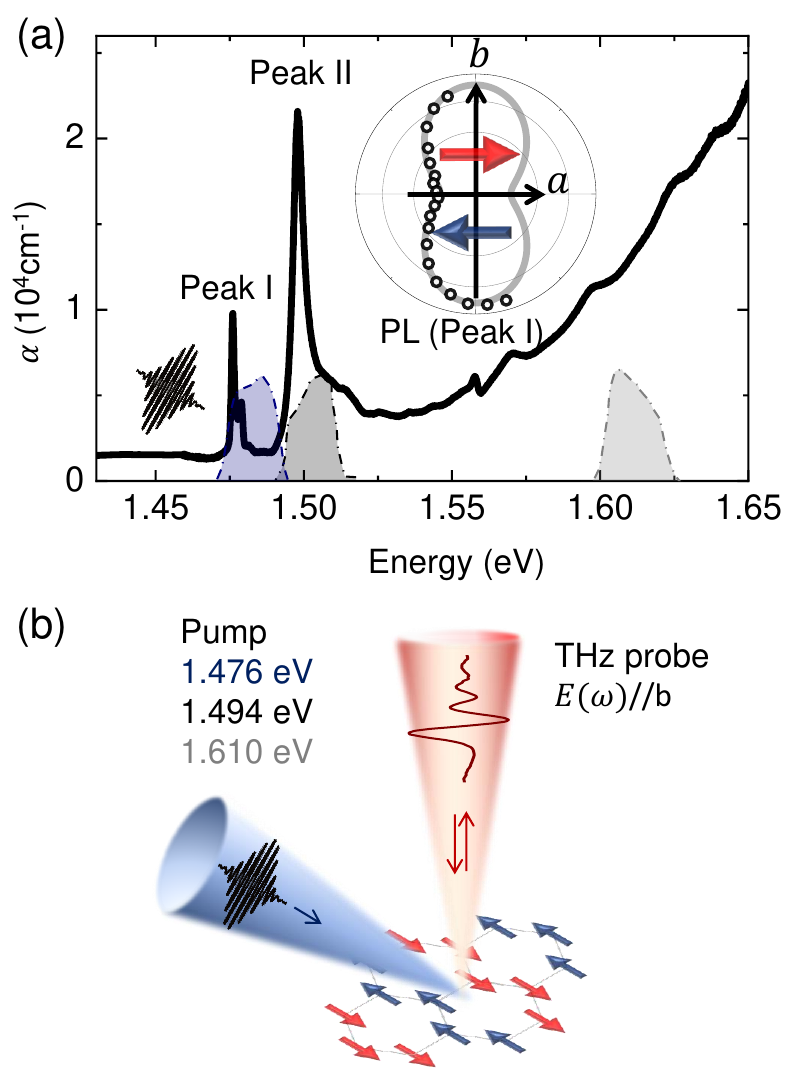}
\caption{ (a) Equilibrium absorption coefficient $\alpha$ at 4 K (data from Ref.\cite{Kang2020}), and three excitation spectra used in this study (shaded peaks). (Inset) The PL intensity maximizes along the $b$ axis, where the spins on the Ni lattice are antiferromagnetically ordered. (b) The transient state was probed by THz pulses using normal incidence reflection.}{\label{fig:PL} }
\end{figure}

%%%%%%%%%%%%%%%%
%%% Results %%%
%%%%%%%%%%%%%%%% 
Millimeter-thick bulk single crystals of NiPS$_3$ with a N\'eel temperature ($T_N$) of 155 K were grown by the chemical vapor transport method\cite{Wang2021}. Within individual NiPS$_3$ layers, the spins on the Ni lattice are antiferromagnetically ordered along the $b$ axis and form a zigzag pattern along the $a$ axis. Highly anisotropic exciton features develop in the antiferromagnetic state\cite{Kang2020,Wang2021,Hwangbo2021,Kim2023}. Peak I, seen at 1.476 eV in the optical absorption (Figure \ref{fig:PL}(a)), is a spin-orbit-entangled exciton\cite{Kang2020,Klaproth2023}. This exciton, probed by the PL, is highly polarized with the largest signal along the $b$ axis\cite{Wang2021}, where the spins are antiferromagnetically ordered (Figure \ref{fig:PL}(a) inset). The second sharp optical transition (peak II in Fig.\ref{fig:PL}(a)) is 22 meV above peak I and is identified as a magnon sideband of the exciton peak I\cite{Kang2020,Kaneko}. 

%%%%%%%%%%%%%%%%%%%%%%%%%%
%%% Pump-probe techique%%%
%%%%%%%%%%%%%%%%%%%%%%%%%% 
Optical excitation at three selected frequencies was used to drive NiPS$_3$ out of equilibrium: (1) resonant pumping of the exciton at 1.476 eV, (2) resonant pumping of the magnon sideband of the exciton at 1.494 eV, and (3) off-resonant pumping at 1.61 eV which is slightly above the absorption onset\cite{Ho2021} at 7 K. The shaded peaks in Fig. \ref{fig:PL}(a) are the spectra of three excitation energies used in this study. The pump beam polarization was parallel to the $a$ axis of NiPS$_3$. We performed time-resolved THz spectroscopy to probe the dynamics of the driven state. The THz probe reached the sample at normal incidence in reflection geometry (Fig.\ref{fig:PL}(b)). The probe polarization was set parallel to the $b$ axis of NiPS$_3$. Further details on the pump-probe measurement can be found in the Supplemental Material \cite{Supp}. 

%%%%%%%%%%%%%%%%%%%%%%%
%%% Overlap signal  %%%
%%%%%%%%%%%%%%%%%%%%%%%
\begin{figure*} 
\includegraphics[scale = 0.65]{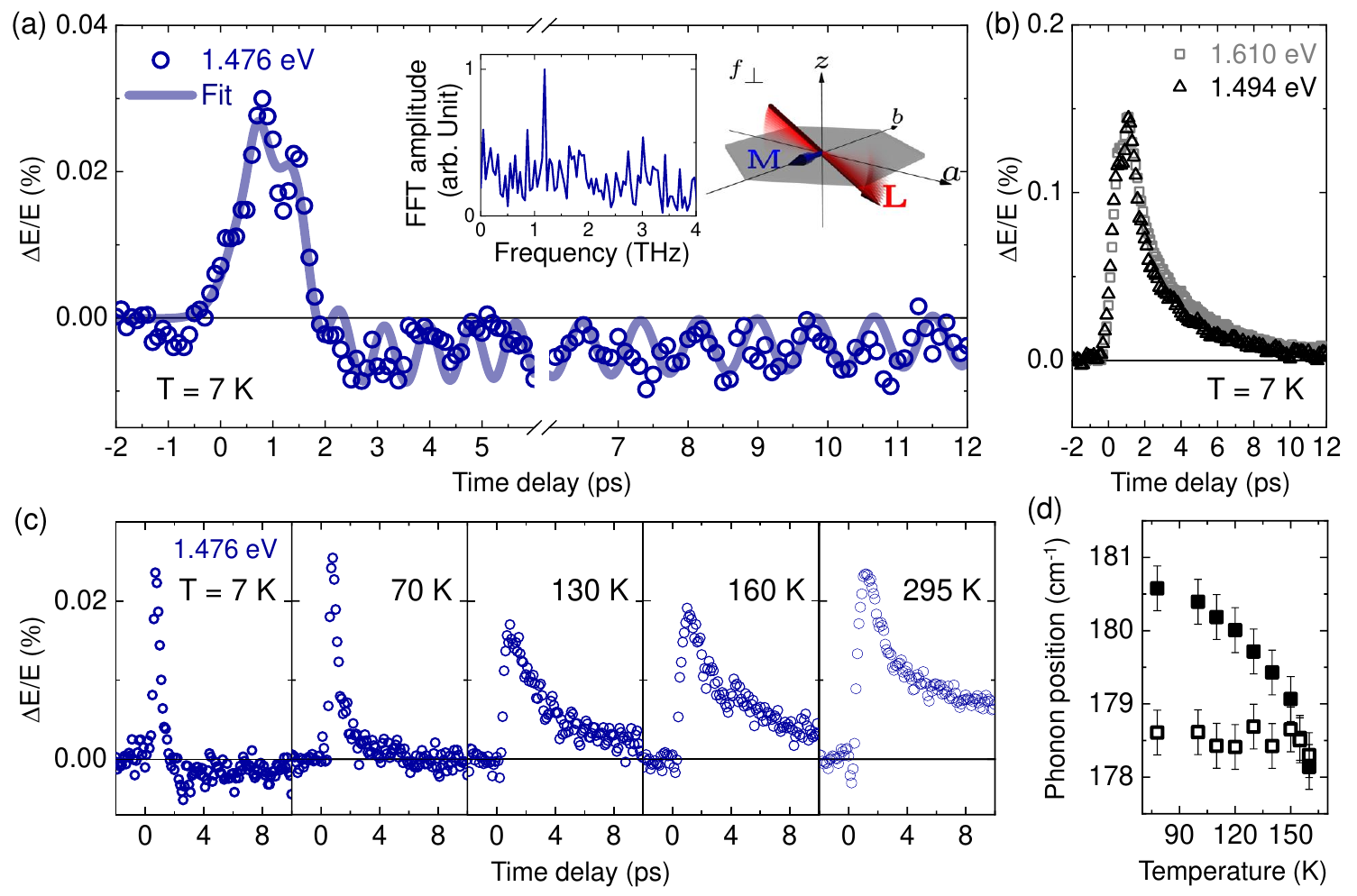}
\caption{(a) Pump-induced change in the THz electric field ($\Delta E/E$) with resonant pumping at peak I at 7 K. $\Delta E/E$ quickly switches sign and remains negative for several tens of picosecond. Meanwhile, an oscillation with a frequency of 1.2 THz (left inset) is seen, corresponding to a magnon mode (right inset). (b) At 7 K, when the pump frequency is tuned to peak II (1.494 eV, triangles) or to above the absorption onset (1.610 eV, squares), $\Delta E/E$ is short lived and remains positive. (c) Temperature evolution of $\Delta E/E$ with 1.476 eV pumping. (d) N\'eel temperature (155 K) characterized by the splitting of a 180-cm$^{-1}$ Raman phonon along the parallel (solid squares) and cross (open squares) polarization geometry\cite{Kim2019}.}{\label{fig:DEoverE}}
\end{figure*}

We first present the spectrally-integrated transient dynamics. Figure \ref{fig:DEoverE}(a) shows the time-domain pump-probe response at 7 K with resonant excitation at peak I. Here, $\Delta E$ is the pump-induced change in the reflected THz field at the peak value, and $E$ is the peak value of the equilibrium THz waveform. After excitation, the pump-probe signal reaches maximum and then decreases to negative values, remaining negative for several tens of picoseconds. Fitting over an extended time window (Fig. S3) reveals two time constants of the transient response. The sign-changing response is short-lived with a lifetime of 0.2 ps, which is likely the time of exciton formation\cite{Steinleitner2017}. The long-lived negative signal has a lifetime of 17 ps, which is on the same order as the lifetime of the exciton\cite{Wang2021,Hwangbo2021,Li2024}. In addition, an oscillatory response is seen with a frequency of 1.2 THz (5.0 meV). This corresponds to an out-of-plane magnon according to the solutions to the Landau–Lifshitz–Gilbert equations (Supplemental Material section S7 \cite{Supp}) and agrees with the magnon frequency observed by Raman scattering\cite{Jana} and THz emission\cite{Belvin2021,Allington2024}. The lifetime of the magnon signal is 70 ps according to our calculation, which fits the THz time trace (Fig. \ref{fig:DEoverE}(a) and Fig. S3).

The negative pump-probe response with a long lifetime is unique to resonant pumping at peak I. As illustrated in Fig. \ref{fig:DEoverE}(b), when pumping at peak II (1.494 eV) or above the absorption edge (1.610 eV) at the same base temperature of 7 K, the pump-probe signals are nearly identical. Only a positive response is seen, and the dynamics are much shorter lived with a lifetime on the order of 3 ps for both cases. A positive $\Delta E/E$ is not surprising for optically excited insulator, as the pump photons above the absorption onset can create hot carriers which enhance material's reflectivity at THz frequencies. In the antiferromagnetic state, pumping away from peak I also leads to a significant reduction in the magnon signature, similar to case of antiferromagnetic NiO\cite{Bossini2021}. 

We then demonstrate the temperature evolution of the pump-probe dynamics. Figure \ref{fig:DEoverE}(c) shows a set of pump-probe overlap scans at various temperatures across $T_N$ = 155 K (see Fig. \ref{fig:DEoverE}(d)) with the same excitation energy of 1.476 eV. With increasing temperature, the exciton peak red-shifts and broadens considerably at 70 K, before nearly disappearing at 130 K\cite{Kang2020,Wang2021,Hwangbo2021,Kim2023}. Accordingly, the suppression in $\Delta E/E$ significantly weakens from 7 K to 70 K, and it is barely recognizable at 130 K. Additionally, the absorption onset also redshifts continuously with increasing temperature so that at room temperature 1.476 eV is above the absorption onset\cite{Ho2021}. Consistently, a positive $\Delta E/E$ is seen at room temperature with 1.476 eV pumping, similar to the off-resonant case at 7 K (Fig. \ref{fig:DEoverE}(b)).

%%%%%%%%%%%%%%%%%%%%%%%%%%%
%%% Frequency-resolved  %%%
%%%%%%%%%%%%%%%%%%%%%%%%%%%
\begin{figure*} 
\includegraphics[scale = 0.9]{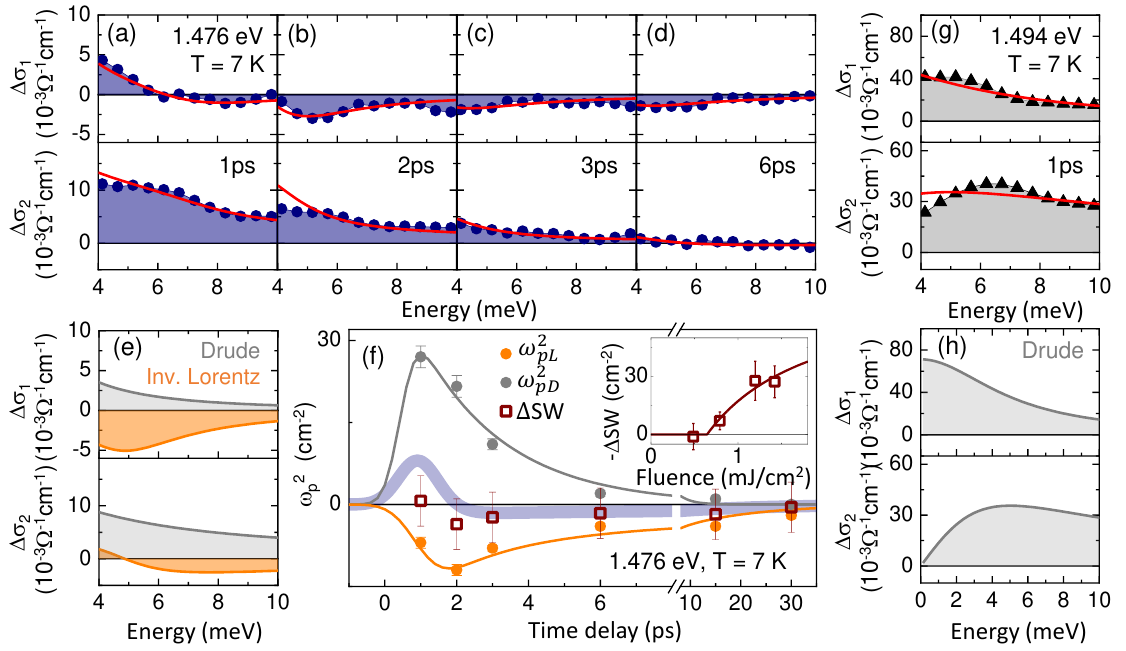}
\caption{(a-d) Resonant pumping at the exciton energy (peak I) results in a long-lived transient suppression in $\sigma_1$ and an increase in $\sigma_2$. The solid circles present the data, and the red curves are fits. (e) The Drude and inverted Lorentz terms used to fit the 2 ps data. (f) Time dependence of the fitting parameters at 7 K with 1.476 eV pumping. $\omega_{pL}^2$ and $\omega_{pD}^2$ represent the oscillator strength of the inverted Lorentz term and the Drude term, respectively. The thin solid lines are exponential decay fits. A global $\omega_p^2$ represented by  the optical spectral weight under $\Delta\sigma_1(\omega)$ in between 4 and 10 meV ($\Delta\text{SW}$). The thick blue line is a rescaled exponential fit used for $\Delta E/E$. (Inset) Transient change in spectral weight as a function of incident pump fluence at $t=3 ps$. (g) Pumping at peak II results in positive $\Delta\sigma_1$ and $\Delta\sigma_2$ (solid triangles) which can be fit with a single Drude term (red lines). (h) The Drude term used to fit \textbf{g} across a broader spectral region. }{\label{fig:Ds1Ds2}}
\end{figure*}
We now investigate the unique long-lived state by the frequency-resolved optical response. Figure \ref{fig:Ds1Ds2}(a-d) shows the time evolution of the light-induced change in the complex optical conductivity, defined as $\Delta\sigma(\omega) = \sigma(\omega)^{\text{transient}} - \sigma(\omega)^{\text{equilibrium}}$. At the maximum response (Fig. \ref{fig:Ds1Ds2}(a)), $\Delta\sigma_1(\omega)$ is positive at low energies and becomes negative in between 6 and 10 meV, while $\Delta\sigma_2(\omega)$ is positive over the entire energy window. At later time delays, $\Delta\sigma_1(\omega)$ remains negative from 4 to 10 meV, while $\Delta\sigma_2(\omega)$ remains positive (Fig. \ref{fig:Ds1Ds2}(b-d)). The real and imaginary part of optical conductivity can be fit simultaneously with a Drude term plus a Lorentz component with a negative amplitude. A typical fit is shown in Fig. \ref{fig:Ds1Ds2}(e). The Drude term gives a positive contribution to $\Delta\sigma_1(\omega)$ for all frequencies, and the inverted Lorentz term brings $\sigma_1(\omega)$ to negative values. Varying the pump-probe time delay, the oscillator strengths for the Drude and inverted Lorentz terms change with different rates, while their widths remain constant ($\gamma^{\text{Drude}}$=1.5 meV and $\gamma^{\text{Lorentz}}$=5 meV). Figure \ref{fig:Ds1Ds2}(f) plots the time evolution of the Drude and Lorentz oscillator strengths. The Drude oscillator strength $\omega_{pD}^2$ can be fit with a single exponential decay with a time constant of 2.4 ps. The inverted Lorentz oscillator strength $\omega_{pL}^2$ reaches maximum at 2 ps then decays with 2 ps and 15 ps time constants. The transient spectral weight, defined as $\Delta SW=\frac{120}{\pi}\int_{\omega_1}^{\omega_2}\Delta\sigma_1(\omega) \,d\omega$, in the range between 4 and 10 meV (open squares in Fig. \ref{fig:Ds1Ds2}(f)) qualitatively follows the exponential fit for the spectrally-integrated response presented in Fig. \ref{fig:DEoverE}(a). A suppression in spectral weight as a result of the negative $\Delta\sigma_1$ is seen up to 30 ps. The frequency-resolved optical response verifies that the overall pump-probe dynamics comes from two contributions with different time scales. We are particularly interested in the inverted Lorentz term, which is a non-thermal response.

This unique non-thermal response disappears when pumping away from peak I. Figure \ref{fig:Ds1Ds2}(g) shows the transient change of the complex optical conductivity with pumping at peak II, which is about 20 meV above peak I. In this configuration, $\Delta\sigma_1(\omega)$ remains positive over the entire energy window, decreasing at higher probe energies, while $\Delta\sigma_2(\omega)$ peaks near 7 meV. This is characteristic of a free-carrier Drude response (Fig. \ref{fig:Ds1Ds2}(h))\cite{Dressel2002}. Similar Drude behavior is seen for 1.476 eV pumping at room temperature (Figure S5) where the exciton disappears and the pump energy is above the absorption edge. These frequency-resolved responses agree with the positive $\Delta E/E$ dynamics as shown in Fig. \ref{fig:DEoverE}(b), thereby verifying the high sensitivity of the transient state to the excitation photon energy: as soon as the excitation energy is at or above the absorption onset, the Drude response dominates.

%%%%%%%%%%%%%%%%%%
%%% Discussion %%%
%%%%%%%%%%%%%%%%%%
We now explore possible origins of the long-lived state with resonant pumping at peak I. As the negative $\Delta\sigma_1(\omega)$ is only observed with the resonant excitation of peak I, and the lifetime of this state is comparable to that of the exciton lifetime\cite{Wang2021,Hwangbo2021,Li2024}, this response must come from optical population of the 1.476 eV exciton level. A negative $\Delta\sigma_1(\omega)$ has been observed in other exciton systems and was interpreted as stimulated emission from population inversion of intraexcitonic levels\cite{Huber2006}. At first, a population inversion is formed by resonant excitation by the optical pump, and then the THz probe photons stimulate a coherent emission similar to the process of lasing. In our case, the reflected THz peak field (Fig. \ref{fig:DEoverE}(a)) encompasses multiple contributions including a co-existent Drude term. We therefore focus on the negative $\Delta\sigma_1(\omega)$ in the following discussion. 

The population inversion picture is further verified by the incident pump fluence dependence of the negative spectral weight, $-\Delta$SW, which follows a threshold-saturation behavior expected for population inversion between excited energy levels, as shown in the inset of Fig. \ref{fig:Ds1Ds2}(f) and Supplemental Material section S6 \cite{Supp}. As each pump spectrum covers a relatively broad frequency range with respect to the narrow exciton linewidth (see Fig. \ref{fig:PL}(a)), the exciton state (peak I) can be either the lower or higher level of the population inversion state. Multiple exciton states close to peak I have been predicted by first-principles calculations\cite{Lane2022,Hamad2024}. Additionally, shoulder peaks at closely spaced energies above the 1.476 eV peak have been observed in photoluminescence and absorption measurements\cite{Kang2020}. Therefore, our observation reflects a population inversion involving the 1.476 eV exciton and another nearby exciton level. Note that the inverted Lorentz term in $\Delta\sigma_1(\omega)$ peaks at 5 meV, overlapping with the magnon energy as shown in Fig. \ref{fig:DEoverE}(a) inset. One may speculate that the negative $\Delta\sigma_1(\omega)$ comes from a magnon-facilitated THz emission; however, this is not likely the case because the suppression in $\Delta\sigma_1(\omega)$ has a much broader energy width than the magnon feature\cite{Belvin2021}.

We briefly comment on other possibilities which can lead to a negative $\Delta\sigma_1(\omega)$. In metallic systems such as graphene, optical excitation leads to a dominant increase in the scattering rate of the existing carriers\cite{Jnawali2013, Frenzel2014}. Here, NiPS$_3$ is insulating with no free-carrier response at equilibrium. Similarly, we can rule out trion formation as the cause of the negative $\Delta\sigma_1(\omega)$. Trions arise from the pre-existing free carriers bound with the photo-generated excitons\cite{Liu2014}, but there are no pre-existing free carriers in our case. Finally, although a reduction of the Lorentz oscillator strength of the absorption near 2.2 eV\cite{Kim2018} can result in a negative $\Delta\sigma_1(\omega)$ and a positive $\Delta\sigma_2(\omega)$ in the THz range, no significant changes in the reflectivity near 2.2 eV was observed under similar pumping conditions\cite{Ergeçen2022}. Further pump-probe studies are needed to investigate the dynamics of the multiple excitonic features near 1.5 eV under resonant excitations.

%%%%%%%%%%%%%%%%
%%% Summary %%%
%%%%%%%%%%%%%%%% 
In conclusion, we discover a long-lived light-induced state with negative photoconductivity in bulk NiPS$_3$ by resonant pumping at the exciton energy. We interpret this as a population inversion state involving the excitonic ground state. Our study provides new insights to understand the nature of the spin-orbit-entangled exciton and the fine optical features nearby, for which limited data and diverse interpretations exist\cite{Kang2020,Wang2021,Klaproth2023,Jana,Kim2023,He2024,Hamad2024}. The long-lived state is unique to resonant pumping at the exciton level, while increasing the pump photon energy by about 20 meV results in short-lived free-carrier response. As the narrow-band optical excitations used in this study are approachable for a wide range of table-top time-resolved techniques and large scale facilities such as femtosecond X-rays from free-electron lasers, our findings open up broad research opportunities to explore long-lived excitonic phases in optically driven nonequilibrium states. In particular, the long lifetime of the spin-orbit-entangled exciton and the exciton-magnon coupling in NiPS$_3$ are appealing for exploring exciton dynamics, exciton condensation and exciton-magnon interactions in light-driven states. Furthermore, since the exciton linewidth in NiPS$_3$ remains narrow down to the trilayer level\cite{Wang2021}, atomically-thin NiPS$_3$ can serve as a building block for van der Waals heterojunctions or superlattices to potentially achieve long-lived lasing at terahertz frequencies in reduced dimensions.

This material is based upon work supported by the National Science Foundation under Grant No. 1944957. H. C., D. A. B. L. and W. H. acknowledge support from the U.S. Department of Energy, Office of Science, Office of Basic Energy Sciences Early Career Research Program under Award Number DE-SC-0021305. Work by Q. T., J. T. and X. L. were supported by the National Science Foundation (NSF) under Grant No. (1945364) and the U.S. Department of Energy, Office of Science, Basic Energy Sciences under Award DE-SC0021064. Q.T. acknowledges support of the Laursen Graduate Research Award. We acknowledge the helpful discussions with Matteo Mitrano, Yue Cao, Gregory Fiete and Martin Eckstein. We thank Boston University Photonics Center for technical support. 

%\bibliography{NPS}% Produces the bibliography via BibTeX.

\end{document}